\documentclass[aps,amssymb,prb,twocolumn,showpacs,amsmath]{revtex4-1} \fussy

\usepackage{times}
\tighten

\usepackage{graphicx,amssymb,amsmath} 
\usepackage{bm}

\newcommand{\figwidth}{0.75\textwidth} 

\begin{document}
\title{Magneto-electric point scattering theory for metamaterial scatterers}
\author{Ivana
Sersic}\email{i.sersic@amolf.nl}\homepage{http://www.amolf.nl/research/resonant-nanophotonics/}
\affiliation{Center for Nanophotonics, FOM Institute for Atomic and
Molecular Physics (AMOLF), Science Park 104, 1098 XG Amsterdam, The
Netherlands}
\author{Christelle Tuambilangana}
\author{Tobias Kampfrath}\email{Present address: Fritz-Haber-Institut der Max-Planck-Gesellschaft, Arnimallee 14, 14195 Berlin, Germany}
\author{A. Femius Koenderink}
\affiliation{Center for Nanophotonics, FOM Institute for Atomic and
Molecular Physics (AMOLF), Science Park 104, 1098 XG Amsterdam, The
Netherlands}
\begin{abstract}
We present a new, fully analytical point scattering model  which can
be applied to arbitrary anisotropic magneto-electric dipole
scatterers, including split ring resonators (SRRs), chiral and
anisotropic plasmonic scatterers. We have taken proper account of
reciprocity and radiation damping for electric and magnetic
scatterers with any general polarizability tensor.  Specifically, we
show how reciprocity and energy balance puts constraints on the
electrodynamic responses arbitrary scatterers can have to light. Our
theory sheds new light on the magnitude of cross sections for
scattering and extinction, and for instance on the emergence of
structural chirality in the optical response of geometrically
non-chiral scatterers like SRRs. We apply the model to SRRs and
discuss how to extract individual components of the polarizability
matrix and extinction cross sections. Finally, we show that our
model describes well the extinction of stereo-dimers of split rings,
while providing new insights in the underlying coupling mechanisms.
\end{abstract}
\maketitle

\section{Introduction}
Research in the field of metamaterials is driven by the possibility
to control the properties of light on the nanoscale by using coupled
resonant nanoscatterers to create optical materials with very
unusual effective medium parameters. Engineering arbitrary values
for the effective permittivity $\epsilon$ and permeability $\mu$
 would allow  new forms of light control based on achieving negative index
materials~\cite{Veselago68,Pendry00,Pendry01}, or transformation
optics media~\cite{Pendry06} that arbitrarily reroute light through
space. In order to reach such control over $\epsilon$ and $\mu$,
many metamaterial building nanoblocks have previously been
identified as having an electric and magnetic response to incident
light, including split ring resonators
(SRRs)~\cite{Smith00,Enkrich05,Rockstuhl06,Klein06,Sersic09,Lahiri10},
rod-pairs~\cite{Shalaev05}, cut-wire pairs~\cite{Dolling05}, fishnet
structures~\cite{Dolling06, Dolling07,Valentine08} and coaxial
waveguides~\cite{Waele10}. In many instances, the nanoscatterers are
not only interesting as building blocks in subwavelength lattices of
designed $\epsilon$ and $\mu$. The building blocks are in fact very
strong scatterers with large cross
sections~\cite{Husnik08,Rockstuhl06b,Corrigan08,Pors10}, comparable
to the large cross sections of plasmonic structures. Therefore,
metamaterial building blocks are excellently suited to construct
magnetic antennas, array waveguides and gratings in which electric
and magnetic dipoles couple and form cooperative excitations, in
analogy to the functionality imparted by plasmon
hybridization~\cite{NordlanderScience}. Experiments outside
the domain of effective media have appeared only recently. These
experiments include experiments by Husnik \emph{et
al}.~\cite{Husnik08}, and Banzer \emph{et al}.~\cite{Banzer10} that
quantify the extinction cross section of single split rings under
differently polarized illumination, as well as a suite of
experiments on coupled systems. These experiments include extinction
measurements on split ring dimers~\cite{Feth10} that point at
resonance hybridization, as well as  reports of magnetization
waves~\cite{decker09}, structural and geometrical chirality in
arrays, as evident in e.g. massive circular
dichroism~\cite{Gansel09,Plum09,Plum09b,Zhang09,Wang09,plum07,decker07,decker10},
and chiral effects in split ring stereo-dimers studied by Liu
\emph{et al}.~\cite{Giessen09}.

In order to understand the light-metamatter interaction in systems
of strongly coupled magneto-electric scatterers, it is important to
understand how individual metamaterial building blocks are excited
and how they scatter.    So far, explanations of the observed
phenomena have mainly rested on two pillars. On the one hand, data
are compared to brute force finite-difference time-domain (FDTD)
simulations of Maxwell's equations, usually showing good
correspondence~\cite{Smith00,Enkrich05,Rockstuhl06b,Klein06,Husnik08,Gansel09,Plum09}.
While the FDTD method is a rigorous method, such numerical experiments do not
in themselves provide insight into how split rings scatter or hybridize in coupled systems. There is general
consensus that to lowest order, metamaterial interactions  in
lattices of scatterers like SRRs must be described by
magneto-electric point-dipole interactions. Hence, simple LC circuit
models with dipolar coupling terms are the second main
interpretative tool to predict, e.g., frequency shifts due to
electric and magnetic dipole-dipole interactions in lattices and
oligomers~\cite{Sersic09,Giessen09,Feth10,guo07}. To rationalize
this LC circuit intuition, several authors have analyzed current
distributions obtained by FDTD simulations in order to retrieve the
microscopic parameters (\emph{i.e.}, the polarizability) underlying
such a dipolar interaction model, and in order to estimate
multipolar
corrections~\cite{Giessen10,Rockstuhl06b,Rockstuhl07,Zhou07,Corrigan08,Pors10,RockstuhlMulti}.

  While there is general consensus that to lowest order,
metamaterial interactions must essentially be magneto-electric point
dipole interactions, we note that  the dipolar circuit models in use
so far have some significant shortcomings. Strictly speaking, the electric circuit
theories lack the velocity of light $c$ as a parameter. Hence, they
contain no retardation or interference, they violate the optical
theorem, do not predict quantitative cross sections, and fail to
predict the effects of super- and subradiant damping on resonance
linewidths. A fair comparison of intuitive point-dipole ideas with
actual data is therefore impossible, unless a fully electrodynamic
theory for magneto-electric point dipoles is derived. Such a theory
would generalize the electric point scattering theory that is well
known as very effective means to describe random media,
extraordinary transmission and plasmon particle
arrays~\cite{Lagendijk96,Pedro98,Abajo07}. In this paper we derive
exactly such a theory for general magneto-electric scatterers.
 We show how
reciprocity and energy conservation restrict the full
magneto-electric response via Onsager
constraints~\cite{Landau,Lindell_biisotropicbook}, and a new
magneto-electric optical theorem for the full polarizability tensor.
This tensor not only includes an electric (magnetic) response to
electric (magnetic) driving, but also off-diagonal coupling in which
a magnetic (electric) response results from electric (magnetic)
driving. While our theory sheds no light on the microscopic origin
of the polarizability~\cite{MerlinPNAS}, we show how electrodynamic
polarizability tensors can be directly constructed from LC circuit
models.  Furthermore we predict how extinction measurements and
measurements of radiation patterns (i.e., differential scattering
cross section) can be used to quantify the polarizability tensor.

 The paper is
structured in the following way: Firstly, in
Section~\ref{section:generaltheory} we derive the general theory,
taking into full account reciprocity, the optical theorem and
radiation damping. In Section~\ref{section:singleSRR} we apply this
theory to set up the polarizability of the archetypical metamaterial
building block, a single  SRR. In Section~\ref{section:singleSRRexp}
we show which set of experiments can be used to retrieve the tensor
polarizability $\boldsymbol{\alpha}$. We find
that magneto-electric coupling directly implies circular dichroism
in the extinction of single split rings, evidencing the utility of
our theory to describe structural chirality~\cite{Gansel09,Plum09,Plum09b,Zhang09,Wang09,plum07,decker07,decker10}. Thirdly,
we show in Section~\ref{section:coupledSRR} that the theory can be
simply applied to obtain quantitative scattering spectra of coupled
systems. By way of example we examine the case of two coupled
resonators in the stereodimer configuration reported by Liu et
al.~\cite{Giessen09}.

\section{Magneto-electric point scatterer\label{section:generaltheory}}
\subsection{Polarizability}
A paradigm in scattering theory is the point dipole
scatterer~\cite{Lagendijk96,Pedro98,Weber04,Koenderink06,Abajo07}
to model scattering by very small, but strongly scattering
particles. Generally, incident fields $\bm{E}$ and $\bm{H}$ induce a (complex) current distribution
in an arbitrary scatterer.
It is the express point of this paper to assess what the scattering properties are of subwavelength scatterers with strong electric and magnetic dipole
moments, as this represents the physics expected of metamaterial building blocks~\cite{Garcia-Garcia05,Sersic09,Giessen09,Feth10,guo07}. Therefore we retain
only electric and magnetic dipole terms, neglecting higher order multipoles.
In such a theory, each scatterer is approximated as an
electric dipole with an electric dipole moment
$\bm{p}=\alpha_{EE}\bm{E}$ that is proportional to the driving
electric field $\bm{E}$. The proportionality constant is the
polarizability $\alpha_{EE}$.
 In this paper, we
derive a generalized point scattering theory for metamaterials that
includes a magnetic dipole moment $\bm{m}$ on an equal footing with
the electric dipole moment $\bm{p}$. In the most general case, the
electric dipole moment $\bm{p}$ and magnetic dipole moments $\bm{m}$
are induced by both the external electric and magnetic fields
$\bm{E}$ and $\bm{H}$ according to
\begin{equation}
\left(
\begin{array} {c}
\bm{p}\\
\bm{m}
\end{array}\right)
=\boldsymbol{\alpha} \left(
\begin{array} {c}
\bm{E}_{\mathrm{in}}\\
\bm{H}_{\mathrm{in}}
\end{array}\right).
\label{Eq:polarizdef}
\end{equation}
 Throughout this paper we suppress
harmonic time dependence $e^{-i\omega t}$. We use a rationalized unit system that significantly
simplifies all equations and is fully explained in
Appendix~\ref{appendix}. In Eq.~(\ref{Eq:polarizdef}),
$\boldsymbol{\alpha}$ is a 6$\times$6
polarizability tensor,
 which consists of four 3$\times$3 blocks, each of which describes part of
the dipole response to the
 electric or magnetic component of the incident light
 \begin{equation}
 \boldsymbol{\alpha}=
 \left(
 \begin{array}{c c}
 \boldsymbol{\alpha}_{EE} & \boldsymbol{\alpha}_{EH}\\
 \boldsymbol{\alpha}_{HE} &
 \boldsymbol{\alpha}_{HH}
\end{array} \right).
\label{Eq:splitpolariz}
\end{equation}
Here, $\boldsymbol{\alpha}_{EE}$ quantifies the
electric dipole induced by an applied electric field. The tensorial
nature of $\boldsymbol{\alpha}_{EE}$ is well
appreciated in scattering theory for anisotropic particles, such as
plasmonic ellipsoids~\cite{BohrenHuffman}. By analogy with the
electric response to electric driving quantified by $
\boldsymbol{\alpha}_{EE}$, the tensor $
\boldsymbol{\alpha}_{HH}$ quantifies the
magnetic dipole induced by a driving magnetic field. Finally, the
off-diagonal blocks represent magneto-electric coupling. The lower
diagonal $ \boldsymbol{\alpha}_{HE}$ quantifies
the magnetic dipole induced by an incident electric field, and $
\boldsymbol{\alpha}_{EH}$ quantifies the
electric dipole induced by an incident magnetic field. Such
magneto-electric coupling is well known to occur in the constitutive
tensors of  metamaterials~\cite{Rockstuhl06}. Indeed, the first
metamaterials consisted of split ring resonators, in which there is
a magnetic response without any driving magnetic field in normal
incidence experiments~\cite{APLSoukoulis}. However, the relative
strength of magneto-electric coupling in the polarizability, i.e., $
\boldsymbol{\alpha}_{EH}$, and
$\boldsymbol{\alpha}_{HE}$ have not been
experimentally quantified.

\subsection{Electrodynamic Onsager relation}
There are several constraints on
$\boldsymbol{\alpha}$. In addition to any
symmetry of the scatterer itself that may impose zeros in the
polarizability tensor, these constraints are due to reciprocity and
to energy conservation. We start by examining the constraints
imposed by reciprocity. It is well known from the field of
bi-anisotropic materials~\cite{Lindell_biisotropicbook} that
reciprocity imposes so-called Onsager constraints on the most
general constitutive tensors relating ($\bm{D},\bm{B}$) to
($\bm{E},\bm{H}$). Already Garc\'{\i}a-Garc\'{\i}a \emph{et
al.}\cite{Garcia-Garcia05} proposed that such Onsager constraints
carry over directly to electrostatic polarizabilities. Here we
rigorously derive Onsager relations for electrodynamic
magneto-electric point scatterers. By, definition, the electric and
magnetic fields due to the induced $\bm{p}$ and $\bm{m}$ are equal
to
\begin{equation}
\left(
\begin{array}{c}
\bm{E}_{\mathrm{out}}\\
\bm{H}_{\mathrm{out}}
\end{array}\right)
=\boldsymbol{G}^0(\bm{r},\bm{r}')\left(
\begin{array}{c}
\bm{p}\\
\bm{m}
\end{array}\right),
\label{Eq:dipoleG}
\end{equation}
with a dyadic Green tensor $\boldsymbol{G}^0$ that describes the field at
position $\bm{r}=(x,y,z)$ due to a dipole at $\bm{r'}=(x',y',z')$.
The 6$\times$6 Green dyadic of free space can be divided in four
3$\times$3 blocks
\begin{equation}
\boldsymbol{G}^0(\bm{r},\bm{r}') =\left(
\begin{array} {c c}
\boldsymbol{G}^0_{EE}(\bm{r},\bm{r}') & \boldsymbol{G}^0_{EH}(\bm{r},\bm{r}')\\
\boldsymbol{G}^0_{HE}(\bm{r},\bm{r}') & \boldsymbol{G}^0_{HH}(\bm{r},\bm{r}')
\end{array} \right)
\label{Eq:freeG}
\end{equation}
The 3$\times$3 diagonals correspond to the familiar known electric
field Green dyadic~\cite{Pedro98,Abajo07} and magnetic field
Green dyadic of free space, which in our unit system (see Appendix) both equal
\begin{equation}
\boldsymbol{G}^0_{EE}(\bm{r},\bm{r}')=\boldsymbol{G}^0_{HH}(\bm{r},\bm{r}')=
(\mathbb{I}k^2+\nabla\nabla)
 \frac{e^{ik|\bm{r}-\bm{r'}|}}{|{\bm{r}-\bm{r'}}|}.
\label{Eq:freeGEE}
\end{equation}
The off diagonal blocks correspond to the mixed dyadics that specify
the electric field at $\bm{r}$ due to a magnetic dipole at
$\bm{r'}$, respectively the magnetic field at $\bm{r}$ due to an
electric dipole at $\bm{r'}$. Explicitly:
\begin{eqnarray}
\boldsymbol{G}^0_{EH}(\bm{r},\bm{r}') & =&
-\boldsymbol{G}^0_{HE}(\bm{r},\bm{r}') \nonumber \\ & =& ik \left(
\begin{array} {c c c}
0 & \partial_z & -\partial_y\\
-\partial_z & 0 & \partial_x\\
\partial_y & -\partial_x & 0
\end{array} \right)\frac{e^{ik|\bm{r}-\bm{r'}|}}{|{\bm{r}-\bm{r'}}|}.
\nonumber\\ \label{Eq:freeGEH}
\end{eqnarray}
In this work we focus solely on scatterers made from reciprocal
constituents, as is commonly true for the metallic scatterers that
constitute metamaterials. Since the materials that compose our
scatterers (typically gold and silver) satisfy reciprocity microscopically, the polarizability
tensor must also lead to a scattering theory that satisfies
reciprocity.

To derive reciprocity constraints on
$\boldsymbol{\alpha}$, it is sufficient to
examine the Green function in the presence of just one point
scatterer at the origin. This Green function that quantifies the
field at $\bm{r}_2$ due to a source at $\bm{r}_1$ in presence of a
single scatterer at $\bm{r}_s$ can be written as~\cite{Lagendijk96,Pedro98,tfootnote}
\begin{equation}
\boldsymbol{G}(\bm{r}_1,\bm{r}_2)=\boldsymbol{G}^0(\bm{r}_1,\bm{r}_2)+\boldsymbol{G}^0(\bm{r}_2,\bm{r}_s)\boldsymbol{\alpha}\boldsymbol{G}^0(\bm{r}_s,\bm{r}_1),
\label{Eq:reciprocity}
\end{equation}
Reciprocity requires for any Green function $\boldsymbol{G}$ (similarly
split in four blocks) that
\begin{eqnarray}
\left(
\begin{array} {c c}
\boldsymbol{G}_{EE}(\bm{r}_2,\bm{r}_1) & \boldsymbol{G}_{EH}(\bm{r}_2,\bm{r}_1)\\
\boldsymbol{G}_{HE}(\bm{r}_2,\bm{r}_1) & \boldsymbol{G}_{HH}(\bm{r}_2,\bm{r}_1)
\end{array}\right) = \qquad\qquad \nonumber\\
 \qquad\qquad \left(
\begin{array} {c c}
\boldsymbol{G}_{EE}(\bm{r}_1,\bm{r}_2) & -\boldsymbol{G}_{EH}(\bm{r}_1,\bm{r}_2)\\
-\boldsymbol{G}_{HE}(\bm{r}_1,\bm{r}_2) & \boldsymbol{G}_{HH}(\bm{r}_1,\bm{r}_2)
\end{array}\right)^T \nonumber  \\
\label{Eq:reciprocTRANSPOSE}
\end{eqnarray}
which is equivalent to noting that swapping source and detector
leaves the detected field unchanged, up to a change in sign.
Specifically, Lorentz reciprocity requires a transpose for the
diagonal 3$\times$3 blocks, meaning that swapping a detector and
source of like character leaves the detected field unchanged. An
extra minus occurs  for the off-diagonal terms, \emph{i.e.}, when
  swapping a magnetic (electric) detector with
an electric (magnetic) source. It is easy to verify that
Eq.~(\ref{Eq:reciprocTRANSPOSE}) is indeed satisfied by the
free space Green function $\boldsymbol{G}^0$.

Using this fact, we evaluate
Eq.~(\ref{Eq:reciprocTRANSPOSE}) for the Green function in
Eq.~(\ref{Eq:reciprocity}) to find if reciprocity constrains
$\boldsymbol{\alpha}$. Since reciprocity is
clearly satisfied for the first term in Eq.~(\ref{Eq:reciprocity}), we
now focus on the second term
\begin{equation}
\boldsymbol{G}^0(\bm{r}_2,\bm{r}_s)\boldsymbol{\alpha}\boldsymbol{G}^0(\bm{r}_s,\bm{r}_1)=\boldsymbol{G}^0(\bm{r}_1,\bm{r}_s)
\boldsymbol{\alpha}\boldsymbol{G}^0(\bm{r}_s,\bm{r}_2).
\end{equation}
Expanding the matrix products in  Eq.~(\ref{Eq:reciprocTRANSPOSE})
while making use of the reciprocity of the free Green function
results in the Onsager relations for the dynamic polarizability:
\begin{eqnarray}
\boldsymbol{\alpha}_{EE}=\boldsymbol{\alpha}_{EE}^T,
\quad
\boldsymbol{\alpha}_{HH}=\boldsymbol{\alpha}_{HH}^T,
\quad
 \mbox{and} \quad
\boldsymbol{\alpha}_{EH}=-\boldsymbol{\alpha}_{HE}^T
\qquad
\nonumber \\
 \label{Eq:onsager}
\end{eqnarray}
These relations are identical in form to the Onsager relations for
constitutive tensors~\cite{Lindell_biisotropicbook}, but are now
derived on very different grounds  for the polarizability of
electrodynamically consistent point scatterers. This gratifying
result shows that the general point dipoles proposed in this work
can be used as microscopic building blocks for an exact scattering
theory that describes the formation  of bi-anisotropic media from dense
lattices of scatterers in the effective medium limit.
Indeed, since the point scattering building blocks fulfill the
Onsager constraints, they are natural building blocks to derive
effective media constitutive tensors by homogenization that also
satisfy the Onsager relations.

\subsection{Optical theorem}
It is well known in point scattering theory for electric dipoles
that polarizability tensors are not solely limited by reciprocity
and spatial symmetry, but also fundamentally by energy conservation.
Indeed, energy conservation imposes an 'optical theorem' that
constrains the polarizability of an electric dipole scatterer to
ensure that  (in absence of material absorption) extinction equals
scattering~\cite{Pedro98}. We proceed to examine these
constraints imposed on $\boldsymbol{\alpha}$.
Let us first recapitulate the well known case of a scalar electric
scatterer~\cite{Lagendijk96,Pedro98}. An electric scatterer will absorb and
scatter part of the incoming light, that together make up the
extinction of a dipole. Extinction for an electric scatterer
corresponds to the work done by the incident field $\bm{E}$
in order to drive the dipole $\bm{p}$.
The work per optical cycle needed to drive $\bm{p}$ equals
$W= \ll\mathrm{Re}{\bm{E}}\cdot \mathrm{Re}{\frac{d\bm{p}}{dt}}\gg$,
where $\ll\gg$ indicates cycle averaging.  Evaluating the work per
cycle, and dividing it by the incident intensity
$I_{\mathrm{in}}=|\bm{E}|^2/2Z$ ($Z$ the impedance of the host
medium) leads to $\sigma_{\mathrm{ext}}=W/I_\mathrm{in}=4\pi k
\mathrm{Im}\alpha_{EE}$. Scattering corresponds to far field radiation
radiated by the dipole $\bm{p}$. According to Larmor, the
cycle-averaged
scattered power is~\cite{jacksonbook} $
P=\frac{4\pi k^4}{3 Z}|\bm{p}|^2.
$ Hence one obtains the well known cross sections
\begin{equation} \sigma_{\mathrm{ext}}=4\pi k \mathrm{Im}\alpha_{EE}
\quad\mbox{and}\quad
\sigma_{\mathrm{scatt}}=\frac{8\pi}{3}k^4|\alpha_{EE}|^2.\label{Eq:scalarcrossect}
\end{equation}
Equating extinction to scattering for nonabsorbing particles to
impose energy conservation, gives rise to  the optical theorem for
the polarizability
\begin{equation}
\mathrm{Im}\alpha=\frac{2}{3}k^3|\alpha_{EE}|^2
\label{Eq:scalarOpticalTheorem}
\end{equation}
This equation for instance shows the well-known fact that a real (electrostatic) $\alpha_0$, such as
Rayleigh's polarizability $\alpha=3V(\epsilon -1)/(\epsilon +2)$ for
a small sphere of dielectric constant $\epsilon$, never satisfies
the optical theorem~\cite{vdHulst}. An  electrostatic $\alpha_0$ can be made to
satisfy the optical theorem by adding radiation
damping~\cite{Pedro98,Abajo07} to obtain the dynamic polarizability
\begin{equation}
\frac{1}{\alpha}=\frac{1}{\alpha_{0}}-i\frac{2}{3}k^3.
\label{Eq:scalarRadiationDamping}
\end{equation}
It is easy to verify  that the albedo
 of a scatterer with
polarizability given by Eq.~(\ref{Eq:scalarRadiationDamping}) is
$$a=\frac{\sigma_\mathrm{scat}}{\sigma_\mathrm{ext}}=\frac{1}{1+\frac{2}{3}k^3\mathrm{Im}\alpha_0},$$ confirming
 that radiation damping indeed transforms any lossless  electrostatic
 polarizability ($\mathrm{Im} \alpha_0=0$) into a scatterer that satisfies
 the optical theorem. Also material loss included in $\alpha_0$ via $\epsilon$
evidently leads to a lossy scatterer $a<1$, as expected.   Many
alternative derivations of Eq.~(\ref{Eq:scalarRadiationDamping}) have appeared, for instance by
making a size parameter expansion of dipolar Mie
coefficients~\cite{wokaun}.

Inspired by the case of a simple electric dipole, we now generalize
the optical theorem and the concept of radiation damping to the full
6x6 tensorial polarizability of arbitrary magneto-electric
scatterers. In this case, the work done per unit cycle by the
incident field $\bm{E}_\mathrm{in}$ and $\bm{H}_\mathrm{in}$ to
drive $\bm{p}$ and $\bm{m}$ is equal to
\begin{equation}
{W}=\quad \ll\mathrm{Re}\bm{E}_{in}\cdot\mathrm{Re}\frac{d\bm{p}}{dt}+\mathrm{Re}\bm{H}_{in}\cdot\mathrm{Re}\frac{d\bm{m}}{dt}\gg
\end{equation}
which evaluates to
\begin{equation}
{W}= \frac{2\pi}{Z} k \mbox{Im} \left[ \left(
\begin{array} {c c}
\bm{E}_\mathrm{in} & \bm{H}_\mathrm{in}
\end{array}\right)^{*}
\boldsymbol{\alpha} \left(
\begin{array} {c}
\bm{E}_\mathrm{in}\\
\bm{H}_\mathrm{in}
\end{array} \right)\right],
\end{equation}
where $(\cdot)^{*}$ indicates complex conjugate.
The power per solid angle radiated by the induced dipoles  in a
direction specified by a unit vector $\bm{\hat{r}}$ is easily found
by calculating the far-field Poynting vector from
Eq.~(\ref{Eq:dipoleG}). The result is composed of three terms:
\begin{equation}
\frac{dP}{d\Omega}=\frac{dP_p}{d\Omega}+\frac{dP_m}{d\Omega}+\frac{k^4}{2Z}\mathrm{Re}(\bm{p}\times
\bm{m})\cdot {\hat{\bm{r}}},\ \label{Eq:radiationpattern}
\end{equation}
 The first term in Eq.~(\ref{Eq:radiationpattern})
represents the scattered radiation of just the electric dipole
$\bm{p}$, which integrates to a total scattered power given by
Larmor's formula. The second term in Eq.~(\ref{Eq:radiationpattern})
represents the radiation pattern of just the magnetic dipole
$\bm{m}$, again given by Larmor's formula. Note that both terms
simply represent the well known $\sin^2\theta$ donut shaped
radiation pattern for $\bm{p}$ and $\bm{m}$. The third term,
however, can completely change the radiation pattern, as it contains
the interference between the fields of $\bm{p}$ and $\bm{m}$. Hence
the relative phase between the induced $\bm{p}$ and $\bm{m}$ is
important for the differential scattering cross section. To obtain
the total scattered power, one should integrate
Eq.~(\ref{Eq:radiationpattern}) over all solid angle. The interference term integrates to 0, as
is easily seen from the fact it is an odd function of
$\hat{\bm{r}}$. Therefore, Larmor's formula immediately generalizes,
and the scattering cross section equals:
\begin{equation}
P=\frac{4\pi}{3Z}k^4\left\|
\begin{array} {c} \bm{p}\\
\bm{m}\end{array}\right\|^2.\label{Eq:tensorLarmor}
\end{equation}
Equating extinction to scattering  results in a condition that must
be satisfied for any incident field
$(\bm{E}_\mathrm{in},\bm{H}_\mathrm{in})$
 \begin{eqnarray}
\mathrm{Im}\left[\left(
\begin{array} {c c}
\bm{E}_\mathrm{in} & \bm{H}_\mathrm{in}
\end{array} \right)^{*}\boldsymbol{\alpha}\left(
\begin{array} {c}
\bm{E}_\mathrm{in}\\
\bm{H}_\mathrm{in}
\end{array} \right)\right]
  = \qquad\qquad \qquad\qquad& \nonumber \\
 \qquad\qquad\qquad\qquad
\frac{2}{3}k^3\left[\left(
\begin{array} {c c}
\bm{E}_\mathrm{in} & \bm{H}_\mathrm{in}
\end{array}\right)^{*}\boldsymbol{\alpha}^{*T}\boldsymbol{\alpha}\left(
\begin{array}
{c} \bm{E}_\mathrm{in}\\
\bm{H}_\mathrm{in}
\end{array} \right)\right], &
\nonumber \\  \label{Eq:trickyOpticalTheorem}
\end{eqnarray}
Due to the
tensorial character of $\boldsymbol{\alpha}$ it
is not immediately evident how to extract a useful optical theorem
that constrains   just the polarizability tensor
$\boldsymbol{\alpha}$ without reference to any
incident field $(\bm{E}_\mathrm{in},\bm{H}_\mathrm{in})$. In order
to eliminate $(\bm{E}_\mathrm{in},\bm{H}_\mathrm{in})$ we make the
assumption (verified below for split rings) that
$\boldsymbol{\alpha}$ can be diagonalized. We
call the eigenvectors $\bm{v}_i$, and denote the eigenvalues, which we
will refer to as `eigenpolarizabilities', with $A_i$. Expanding the
incident field at the position of the origin in the orthogonal
eigenvectors
\begin{equation}
\left(
\begin{array} {c}
\bm{E}_\mathrm{in}\\
\bm{H}_\mathrm{in}
\end{array} \right)=\sum_i c_i \bm{v}_i,
\end{equation} and with  $
 \boldsymbol{\alpha} \bm{v}_i=A_i
\bm{v}_i$  and $\langle \bm{v}_i|\bm{v}_j\rangle=\delta_{ij}$,
Eq.~(\ref{Eq:trickyOpticalTheorem}) reduces to
\begin{equation}
\frac{2}{3}k^3{\sum_{i=1}^6}|c_i|^2|A_i|^2\geq{\sum_{i=1}^6}|c_i|^2\mathrm{Im}A_i,\label{Eq:almostOpticalTheorem}
\end{equation}
with strict equality for lossless scatterers. Since this equation
must be satisfied for any choice of incident wave (i.e., any
combination of $c_i$), we find a generalized optical theorem for
6$\times$6 polarizability tensors that can be expressed in terms
of the eigenpolarizabilities as
\begin{equation}
\frac{2}{3}k^3|A_i|^2\geq \mathrm{Im} A_i \qquad \forall i=1\ldots
6, \label{Eq:TensorpticalTheorem}
\end{equation}
 again with strict equality for lossless scatterers.
Eq.~(\ref{Eq:TensorpticalTheorem}) implies that the polarizability
tensor represents an energy conserving scatterer, if and only if
each of its 6 eigenpolarizabilities are chosen to satisfy the simple
scalar optical theorem (Eq.~(\ref{Eq:scalarOpticalTheorem})) derived
for electric scatterers. This general optical theorem highlights the
importance of two new quantities: the eigenpolarizabilities, and the
corresponding eigenvectors
 of the point scatterer polarizability.

In Eq.~(\ref{Eq:scalarRadiationDamping}) we reviewed the well-known
addition of radiation damping required to make electrostatic
polarizabilities satisfy the optical theorem. Since metamaterial
scatterers are frequently treated via electrostatic circuit models,
it would be extremely fruitful to generalize this method to general
6$\times$6 electrostatic polarizability tensors. It is now evident,
that we can simply apply the scalar recipe to each
eigenpolarizability separately. An alternative notation for this
method is:
\begin{equation}
\boldsymbol{\alpha}^{-1}=\boldsymbol{\alpha}_0^{-1}-\frac{2}{3}k^3i\mathbb{I}
\label{Eq:TensorRadiationDamping}
\end{equation}
We note that this expression, which is identical to
Eq.~(\ref{Eq:scalarRadiationDamping}) upon replacement of
$1/(\cdot)$ by matrix inversion, provides a unique relation to
translate a magneto-/electrostatic polarizability tensor
$\boldsymbol{\alpha}_0$ derived from RLC
circuit theory, to the corresponding electrodynamic polarizability
that satisfies the optical theorem. We can hence consistently assess
how intuitive ideas based on a microscopic RLC circuit model for electrostatic dipoles lead to
quantitative predictions for extinction, scattering, as well as
resonance hybridization, diffraction and super/sub radiant damping
in coupled systems, such as periodic systems, or arbitrary finite
clusters.

\section{Polarizability of split ring resonators\label{section:singleSRR}}
\begin{figure*}[t]
\includegraphics[width=\textwidth]{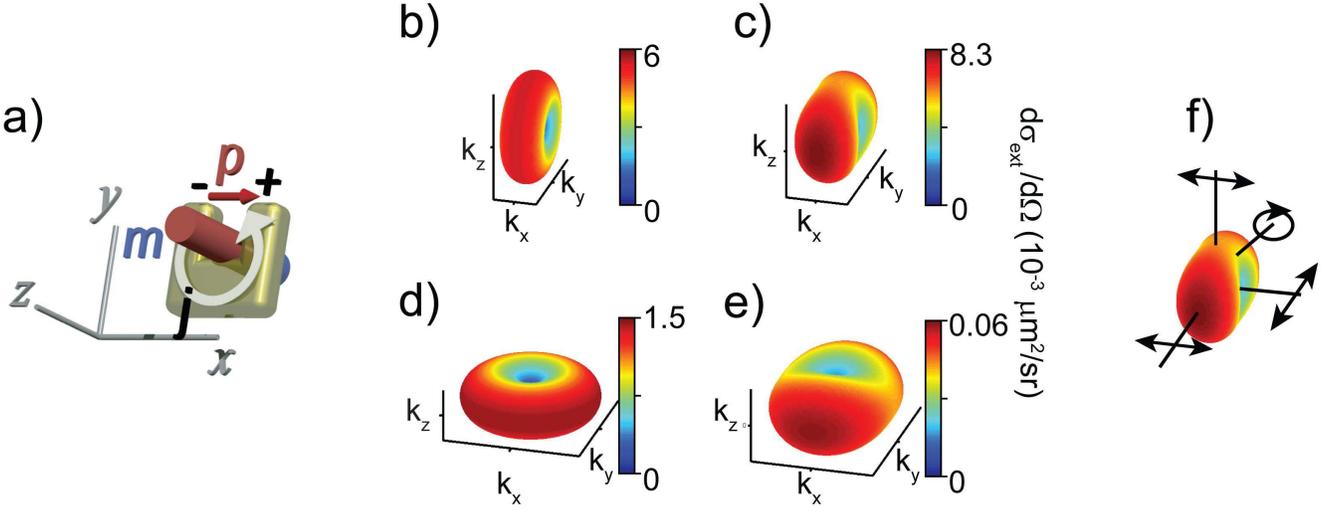}
\caption{Split ring radiation patterns corresponding to the
polarizability tensor eigenvectors. Panel (a): (Sketch) A single
split ring resonator can have an electric dipole moment $\bm{p}$
along the $x$-axis due to charging of the split. Circulating current
$j$ in the ring gives rise to a magnetic dipole moment $\bm{m}$ in
the $z$-direction. Panels (b,c): Radiation patterns of the two
eigenmodes of an SRR in the case of no off-diagonal magneto-electric
coupling ($\eta_E=0.7, \eta_H=0.3, \eta_C=0$). The electric dipole
moment oriented along the $x$-axis radiates most of its amplitude in
the $ky,kz$ plane, while the magnetic dipole oriented along the
$z$-axis radiates mostly into the $kx,ky$ plane. Panels (d,e):
radiation patterns of the eigenvectors with magneto-electric cross
coupling ($\eta_C=0.4$). Panel (f): indication of the polarization
of the light radiated by the eigenvector with largest eigenvalue
(panel (c)).  Light is linearly polarized for wave vectors along the
cartesian axes, but elliptically polarized in general. The direction
of strongest circular dichroism in extinction and scattering is in
the $xz$-plane.} \label{Fig1:radpats}
\end{figure*}
\subsection{Symmetry}
 As an example of our general theory
we consider the specific example of split ring resonators. The
electrostatic polarizability of split ring resonators was
discussed for instance by Garc\`{i}a-Garc\`{i}a et
al~\cite{Garcia-Garcia05}.
 We consider the LC resonance of an infinitely thin split ring
in the $xy$ plane, with split oriented along the $x$ axis, as shown
in Fig.~\ref{Fig1:radpats}(a). Incident electric field polarized
along the $x$ direction gives rise to an electric dipole
$\bm{p}=(\alpha_{EE}^{xx}E_x,0,0)$ oriented along the split of the SRR.
As in an LC circuit, the charge separation generated over the
capacitive split relaxes as a circulating current, hence giving rise
to a magnetic dipole $\bm{m}=(0,0,\alpha_{HE}^{zx}E_x)$ in the $z$
direction, in response to a driving E-field along $x$~\cite{Burresi09}. The same is
valid vice versa: an applied magnetic field along $z$ induces a
magnetic dipole moment $\bm{m}=(0,0,\alpha_{HH}^{zz} H_z)$ along the
$z$ direction. The associated current accumulates at the gap, giving
rise to an electric dipole moment $\bm{p}=(\alpha_{HE}^{xz}H_z,0,0)$
driven by $H_z$. If we assume that the LC resonance really only
involves $\bm{p}_x$ and $\bm{m}_z$, we find that the polarizability tensor
is filled with zeros, except for the four contributions described
above. Hence
\begin{equation} \boldsymbol{\alpha}_{SRR} = \left(
\begin{array} {c c c c c}
\alpha_{EE}^{xx} & 0 & ... & 0 & \alpha_{EH}^{xz}\\
0 &  &  &  & 0\\
\vdots &  & \ddots &  & \vdots\\
0 &  &  &  & 0\\
\alpha_{HE}^{zx} & 0 & ... & 0 & \alpha_{HH}^{zz}
\end{array} \right). \label{Eq:SRRsymmetry}
\end{equation}
The symmetry constraints that set which elements of
$\boldsymbol{\alpha}_{SRR}$ are zero,
are valid  both for the electrodynamic and electrostatic
polarizability of split rings.

\subsection{Quasi-electrostatic RLC model}
We will now construct the electrodynamic polarizability by starting
from an electrostatic polarizability derived from a single resonant
RLC equation of motion.
 Therefore we take a common resonant frequency dependence out of the tensor
elements, writing
 \begin{equation}
\boldsymbol{\alpha}_{SRR}^{\mathrm{static}}=\alpha(\omega)
 \left(
 \begin{array}{c c c c c}
 \eta_{E} & 0 & ... & 0 &  i\eta_{C}\\
 0 &  &  &  & 0\\
\vdots &  & \ddots &  & \vdots\\
0 &  &  &  & 0\\
 -i\eta_{C} & 0 & ... & 0 &\eta_{H}
 \end{array}
\right),\label{Eq:SRRpolarizsimple}
\end{equation}
where $\eta_E$, $\eta_C$ and $\eta_H$ are constant and
$\alpha(\omega)$ is a Lorentzian prefactor
\begin{equation}
\alpha(\omega)=\frac{\omega_0^2
V}{\omega_0^2-\omega^2-i\omega\gamma}. \label{Eq:prefactorSRR}
\end{equation}
Here, $\omega_0$ is the SRR resonance frequency $\omega_0\approx
\frac{1}{\sqrt{LC}}$, $\gamma$ is the damping rate due to the Ohmic
loss of gold and $V$ is the physical particle volume.
As in the plasmonic case, this approximation is coined `quasi-static', as it does contain frequency $\omega$, but does not contain
the velocity of light $c$. The polarizability obtained from the quasi-static polarizability once the radiation damping term is added (section~\ref{magnetoelectricsection} Eq.~(\ref{Eq:TensorRadiationDamping})) is called `dynamic polarizability'.
In this
formulation, all the frequency dependence, and the units of
$\boldsymbol{\alpha}_{SRR}$ are contained in
$\alpha(\omega)$. The parameters $\eta_E$, $\eta_H$ and $\eta_C$ are
dimensionless. For a lossless split ring $\eta_E$, $\eta_H$ and
$\eta_C$ are all real. We assume that all losses are introduced via
$\gamma$. To determine the sign of $\eta_E$, $\eta_H$ and $\eta_C$,
we expect that for very slow driving the charge (current) on the
capacitor directly follows the driving $E$ ($H$)-field, implying
$\eta_E>0$ and $\eta_H>0$. The sign  of $\eta_C$ follows similar
reasoning, After charge build-up, charge associated with a
$\bm{p}_x=\mathrm{Re}(\alpha(\omega) \eta_E e^{-i \omega t} E_x)$
relaxes as counter-clockwise current, giving rise to a negative
$\bm{m}_z=\mathrm{Re}(\alpha(\omega) i \eta_C e^{-i \omega t} E_x)$,
implying that $\mathrm{sign}\,\eta_C=\mathrm{sign}\,\eta_E$.

\subsection{Limit on magneto-electric coupling\label{magnetoelectricsection}}
 Having constructed an
electrostatic polarizability in accordance with RLC circuit
models proposed in earlier reports, we apply radiation damping
according to Eq.~(\ref{Eq:TensorRadiationDamping}) to obtain a
scatterer that has a correct energy balance:
\begin{equation}
\boldsymbol{\alpha}_{{SRR}}^{-1}=(\boldsymbol{\alpha}_{SRR}^{\mathrm{static}})^{-1}-\frac{2}{3}k^3\mathbb{I}.\label{Eq:SRRradiationdamp}
\end{equation}
So far we have not explicitly discussed absorption loss, except
through the inclusion of the material damping constant $\gamma$ in the quasi-static
polarizability. Starting from a quasi-static
polarizability with quasi-static eigenpolarizabilities
$A^{\mathrm{static}}_i$, the albedo for each eigenillumination
$\bm{v}_i$ can be expressed as
\begin{equation}
a_i=\frac{1}{1+\frac{2}{3}k^3\mathrm{Im}A^{\mathrm{static}}_i}.\label{Eq:albedo}
\end{equation}
 It follows that for any lossy scatterer
the imaginary part of each eigenvalue $A^\mathrm{static}_i$ of the
electrostatic polarizability tensor must be positive to ensure
$0\leq a\leq1$. In the case of a tensorial
$\boldsymbol{\alpha}$ with loss included as in
Eq.~(\ref{Eq:SRRpolarizsimple}), ~(\ref{Eq:prefactorSRR}), one needs to
explicitly verify that each eigenvalue has positive imaginary part.
The eigenvalues of Eq.~(\ref{Eq:SRRpolarizsimple})
are$A^\mathrm{static}_\pm=\alpha(\omega)\lambda_\pm$ with
$\lambda_\pm=\frac{\eta_E+\eta_H\pm\sqrt{(\eta_E-\eta_H)^2+4\eta_C^2}}{2}$.
Since $\mathrm{Im}(\alpha(\omega))\geq0$ and $\lambda_{\pm}$ are
real, we find that both eigenvalues have positive imaginary part
only if both $\lambda_+$ and $\lambda_-$ are positive. Thus, loss
sets an additional constraint on the polarizability tensor, and
limits  the magneto-electric coupling to
\begin{equation}
|\eta_{C}|\leq\sqrt{\eta_E\eta_H}. \label{Eq:crosscouplelimit}
\end{equation}
This result implies a very important limitation on magneto-electric
scatterers: it states that a magneto-electric cross coupling
($\eta_C$) can only be generated if there is a sufficiently strong
directly electric, and directly magnetic response. We note that this
constraint is very similar to the constraint on the magneto-electric
cross coupling in constitutive tensors derived for homogeneous
bi-anisotropic media
in Ref.~\onlinecite{Lindell_biisotropicbook} that recently attracted attention in the framework of proposals for
repulsive Casimir forces~\cite{silveirinha,soukouliscasimir}.  While our
derivation was specific for split rings, we note that similar
constraints hold for all magneto-electric scatterers. In the
presence of material loss, the magneto-electric coupling terms are
limited by the fact that all electrostatic eigenpolarizabilities
must have positive imaginary part.

\section{Predicted scattering properties of single split rings\label{section:singleSRRexp}}
 In the
remainder of the paper we discuss some insights that the proposed
magneto-electric point scattering theory provides in how split rings
scatter.  In this section  we will consider the eigenmodes and the
radiation patterns of a single SRR for
 $\boldsymbol{\alpha}$ given by Eq.~(\ref{Eq:SRRradiationdamp}). Next, we predict which
set of experiments will provide full information on the elements of
the polarizability tensor. We will show how the extinction cross
sections can be translated back to retrieve SRR polarizabilities and
magneto-electric cross polarizabilities of a single SRR. Although
the results we present are general, we use a specific set of
parameters for all the figures presented in this paper. These
parameters are chosen to fit to the properties of split rings that
are resonant at $\lambda=1.5~\mu$m ($\omega_0/2\pi= 200$~THz and
that consist of 200~by 200~nm gold split rings with a thickness of
30~nm and a gap width of 90~nm.  Thus we take $V=200\times 200\times
30$~nm$^3$. We set the damping rate to be that of gold $\gamma
=1.25\cdot 10^{14}$~s$^{-1}$ as fitted to optical constants tabulated in in Ref.~\onlinecite{Johnson}. We use $\eta_E=0.7$,
$\eta_H=0.3$ and $\eta_C=0.4$. These parameters were chosen because
(A) they reproduce quantitatively the extinction cross section under
normal incidence along the $z$-axis measured by Husnik \emph{et
al.}~\cite{Husnik08}, and (B) they fit well to our transmission data
on arrays of different densities of split rings taken at normal
incidence~\cite{Sersic09} and as a function of incidence
angle~\cite{Sersictopublish}.   The chosen values correspond to
on-resonance polarizabilities $\alpha_{EE}= 4.6 V$, $\alpha_{HH}=2.1V$ and
$\alpha_{EH}=2.5 V$, all well in excess of the physical SRR volume $V$
as is typical for strong scatterers.  Finally, we note that the  calculated albedo
fits well to the albedo $a=0.5$ to $0.75$ calculated by FDTD by Husnik et al.~\cite{Husnik08}.

\subsection{Radiation patterns and eigenvectors of the polarizability tensor}
 In Fig. ~\ref{Fig1:radpats},
we consider the eigenstates of the split ring polarizability tensor
presented in Eq.~(\ref{Eq:SRRradiationdamp}). We first assume
that the cross coupling terms are absent, i.e., $\eta_C=0$, in which
case the polarizability tensor is diagonal, with
eigenpolarizabilities $\alpha(\omega) \eta_E$ and
$\alpha(\omega)\eta_H$. The corresponding orthogonal eigenmodes are
$(p_x,m_z)=(1,0)$ and $(p_x,m_z)=(0,1)$. Figures ~\ref{Fig1:radpats}
(b) and (c) show radiation patterns of the two eigenmodes.
Figure ~\ref{Fig1:radpats}(b) shows the radiation pattern of the
purely electric eigenmode $(p_x,m_z)=(1,0)$ and Fig.
~\ref{Fig1:radpats}(c) shows the radiation pattern of the purely
magnetic eigenmode $(p_x,m_z)=(0,1)$. Note that both $\bm{p}_x$ and
$\bm{m}_z$ radiate as simple dipoles with a $\sin^2\theta$ far field
radiation pattern~\cite{jacksonbook}. The two eigenmodes can be
selectively excited  by impinging with a plane wave incident along
the $z$-axis with $x$-polarized $E$-field (electric eigenmode), or
with a plane wave incident along the $x$-axis with $y$-polarization
($z$-polarized $H$-field, magnetic eigenmode).  The extinction cross
section of a single split ring at these two incidence conditions is
set by $\sigma_{\mathrm{ext}}=4 \pi k \mathrm{Im}(\alpha_{EE})$ and
$\sigma_{\mathrm{ext}}=4 \pi k \mathrm{Im}(\alpha_{HH})$.

Next, we consider extinction and eigenmodes for arbitrary values of
the cross coupling. It is easy to see that the extinction cross
section at the two special illumination conditions (incident along
$z$, $x$-polarized,  respectively, incident along $x$, with
$y$-polarization) remain equal to $\sigma_{ext}=4 \pi k
\mathrm{Im}(\alpha_{EE})$ and $\sigma_{ext}=4 \pi k
\mathrm{Im}(\alpha_{HH})$.  However, for nonzero $\eta_C$, these
incidence conditions and polarizabilities do not correspond anymore
to the eigenvalues and eigenvectors of the polarizability tensor,
which now have mixed magneto-electric character.  In the extreme
case of strongest magneto-electric coupling
($\eta_C=\sqrt{\eta_E\eta_H}$), the eigenvectors reduce to
$(p_x,m_z)=(1,i\sqrt{\eta_E/\eta_H})$ and
$(p_x,m_z)=(1,-i\sqrt{\eta_H/\eta_E})$.  The associated far-field
radiation patterns of these eigenvectors correspond to coherent
superpositions of the radiation pattern of an $x$-oriented electric
dipole, and a $z$-oriented magnetic dipole, with a quarter wave
phase difference. Figures ~\ref{Fig1:radpats}(d,e) show the
on-resonance radiation pattern, assuming $\eta_E=0.7$, $\eta_H=0.3$
and $\eta_C=$0.4\texttt.
Note that these parameters are close to the limit of strongest
possible magneto-electric coupling. Figures ~\ref{Fig1:radpats}(d,e)
reveal that the radiation pattern of each eigenmode is non-dipolar.
Rather than a $\sin^2\theta$ donut-shaped pattern, an elongated
radiation pattern occurs, with maximum extent in the $y$-direction.
The polarization in the far field is linear for directions
along the cartesian axis, but is generally elliptical.

\subsection{Extinction cross sections to measure
polarizability}
\begin{figure*}
\includegraphics[width=\figwidth]{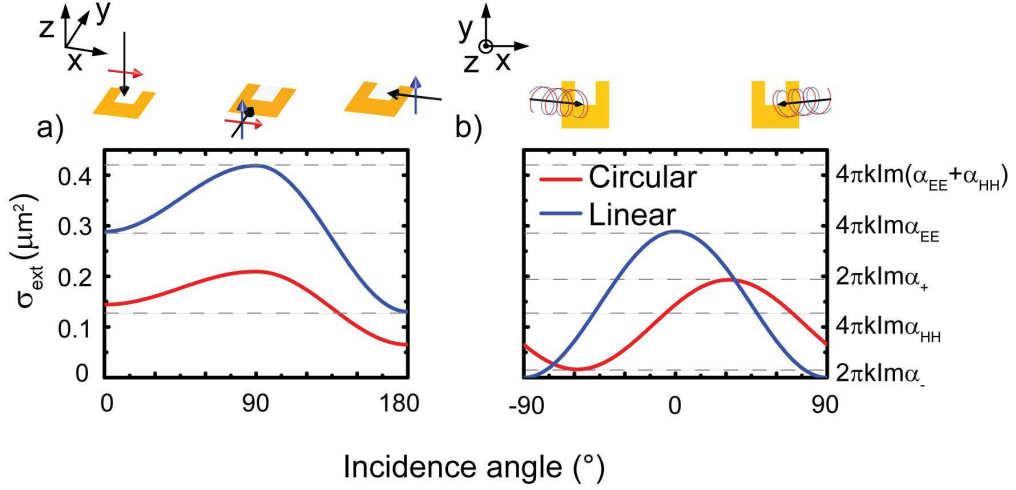}
\caption{Extinction cross section $\sigma_{ext}$ as a function  of
the illumination angle and polarization. Blue lines represent
$\sigma_{ext}$ for linearly polarized incident illumination, while
red lines represent extinction for right handed circularly polarized
illumination. Panel (a) shows extinction for incidence wave vectors
ranging from $k_z$ to $k_y$ to $k_x$. At normal incidence with $k$
along the $z$-axis,  $\sigma_{ext}$ is a measure for only $\alpha_{EE}$
as $E_x$ is the only driving field. Increasing the angle to
$90^\circ$ both polarizations $E_x$ and $H_z$ excite the dipoles in
the SRR, so $\sigma_{ext}$ is a measure for the sum of the terms on
the diagonal of the polarizability tensor $(\alpha_{EE}+\alpha_{HH})$.
Changing the angle to couple only the $H_z$ of the incident light to
the SRR gives $\sigma_{ext}$ that is a measure for purely
$\alpha_{HH}$. Panel (b) $\sigma_{ext}$ as a function of the incident
angle in the $xz$-plane (wave vector ranging from $-k_x$ to $k_z$ to
$k_x$). For right-handed circular polarization minima and maxima in
$\sigma_{ext}$ occur as a function of angle, which are a measure for
the eigenpolarizabilities $\alpha_-$ and $\alpha_+$, respectively.
Both sets of measurements in panel (a) and (b) together provide
information on all the components of the polarizability tensor,
$\alpha_{EE}$, $\alpha_{HH}$, and $\alpha_{EH}$.} \label{Fig2:extinction}
\end{figure*}
Figure~\ref{Fig2:extinction} shows the extinction cross section
predicted by our point scattering model of a single split ring for
different incidence conditions. In Fig.~\ref{Fig2:extinction}(a),
the incident wave vector is swept from the $z$-direction to the
$y$-direction, while maintaining $x-$ polarized light. For this set
of incidence conditions the resulting extinction cross sections only
depend on $\alpha_{EE}$ and $\alpha_{HH}$, and are entirely independent of
the off-diagonal coupling strength $\alpha_{EH}$. The cross section
increases from $\sigma_{\mathrm{ext}}= 4 \pi k\mathrm{Im}\alpha_{EE}$
as the split ring is only driven by the incident $E_x$ field when
light is incident along $z$, to $\sigma_{\mathrm{ext}}= 4 \pi
k(\mathrm{Im}\alpha_{EE}+ \mathrm{Im}\alpha_{HH})$, as the split ring is
driven by the incident $E_x$ field plus the incident $H_z$ field.
When the wavevector is rotated to the the $x$-axis, the extinction
cross section diminishes to $4\pi k \mathrm{Im}\alpha_{HH}$, as the
split ring is only driven by $H_z$. The chosen values $\eta_E=0.7,
\eta_H=0.3$ and $\eta_C=0.4$ that we also used for
Fig.~\ref{Fig1:radpats}(d,e) yield extinction cross sections
$\sigma_{\mathrm{ext}}= 4 \pi k\mathrm{Im}\alpha_{EE}=0.29~\mu$m$^2$
and $\sigma_{\mathrm{ext}}= 4 \pi
k\mathrm{Im}\alpha_{HH}=0.13~\mu$m$^2$. The predicted
$\sigma_{\mathrm{ext}}= 4 \pi k\mathrm{Im}\alpha_{EE}=0.29~\mu$m$^2$ is
consistent with the measurement
($\sigma_{\mathrm{ext}}=0.3~\mu$m$^2$) reported by Husnik et
al.~\cite{Husnik08}. It is  important to note that measurements
along cartesian incidence directions and with linear cartesian
polarizations  yield only the diagonal elements of the
polarizability tensor. Indeed, the proposed measurements form a
redundant set of measurements of $\alpha_{EE}$, $\alpha_{HH}$, and
$(\alpha_{EE}+\alpha_{HH})$, but do not provide any insight into the
magneto-electric cross coupling in the electrodynamic polarizability
tensor.\cite{diagonalnote}

In order to measure the eigenpolarizabilities, it is necessary to
selectively address the eigenvectors of the polarizability tensor.
As noted above, the eigenvectors in the case of strong
magneto-electric coupling $\eta_C\approx\sqrt{\eta_E\eta_H}$ tend
to $(p_x,m_z)= (1,i\sqrt{\eta_E/\eta_H})$ and
$(1,-i\sqrt{\eta_H/\eta_E})$.   These eigenvectors require
simultaneous driving by $E_x$ and $H_z$, with a quarter wave phase
difference. We note that such fields can be generated by
circularly polarized light with incident wave vector constrained
to the $xz$-plane. Indeed, at maximally strong magneto-electric
coupling and $\eta_E=\eta_H$, circularly polarized light  incident
at $45^\circ$ from the $z$-axis would selectively excite exactly one
eigenmode. Therefore, we expect angle-resolved extinction
measurements for oppositely handed circularly polarized beams to reveal the
eigenpolarizabilities. Figure~\ref{Fig2:extinction}(b) plots the
extinction  cross section for right handed circular polarization,
as a function of angle of incidence in the $z$-plane, for
illumination tuned to the LC resonance frequency. Naturally, at
normal incidence the extinction is exactly half the extinction
obtained for linear polarization, as a consequence of the fact
that $\textit{E}_y$ does not interact with the split ring at all.
Strikingly, the extinction cross section is predicted to behave
asymmetrically as a function of incidence angle. The extinction
increases when going to positive angle and decreases when going to
negative angle. Changing handedness is equivalent to swapping
positive and negative angles. A detailed analysis shows that the
maximum in extinction corresponds to the largest eigenvalue of the
polarizability tensor ($\sigma_{ext} = 2 \pi k \mathrm{Im}
\alpha_+$), while the minimum in extinction corresponds to the
smallest eigenvalue ($\sigma_{ext} = 2 \pi k \mathrm{Im}
\alpha_-$). Therefore, circularly polarized measurements reveal
the eigenvalues of the polarizability tensor. Combining such
circularly polarized extinction measurements with the measurements
under cartesian incidence in Fig.~\ref{Fig2:extinction}(a),
therefore allows to extract all components of the polarizability
tensor. In addition to the contrast in  extinction, the angle at
which the maximum circular dichroism occurs is a second,
independent measure for the magneto-electric coupling strength. The
measurements in Fig.~\ref{Fig2:extinction}(a) and (b) together
hence provide full, even redundant, information on $\eta_E$,
$\eta_H$ and $\eta_C$.

\subsection{Structural chirality}
 The results plotted
in Fig.~\ref{Fig2:extinction}(b) show that magneto-electric
coupling in the 6$\times$6 polarizability tensor  directly implies
structural chirality. It is exhilarating that this interesting
phenomenon first reported by~\cite{Plum09,plum07} for the transmission of arrays
of scatterers is naturally present in the theory.  However, while previous analysis of structural chirality focused
on transmission through periodic arrays, we predict that  circular
dichroism already appears in the extinction cross section of a
single split ring, with a strength set by how close the
magneto-electric coupling strength is to its limit $\sqrt{\eta_E,\eta_H}$. The circular dichroism in
extinction occurs independently of whether there is  material loss, as opposed to,
e.g.,  asymmetric transmission phenomena through arrays, that are claimed to
require dissipation~\cite{plum07}. For maximally
magneto-electrically coupled systems,  the smallest eigenvalue is
identically zero, implying that such a scatterer is transparent
for one circular polarization, and achieves its strongest
scattering for the opposite handedness. We expect that our
6$\times$6 polarizability tensor can be successfully used to
describe all structurally chiral scatterers reported today, as
well as clusters and periodic arrays thereof.
\begin{figure*}
\includegraphics[width=\textwidth]{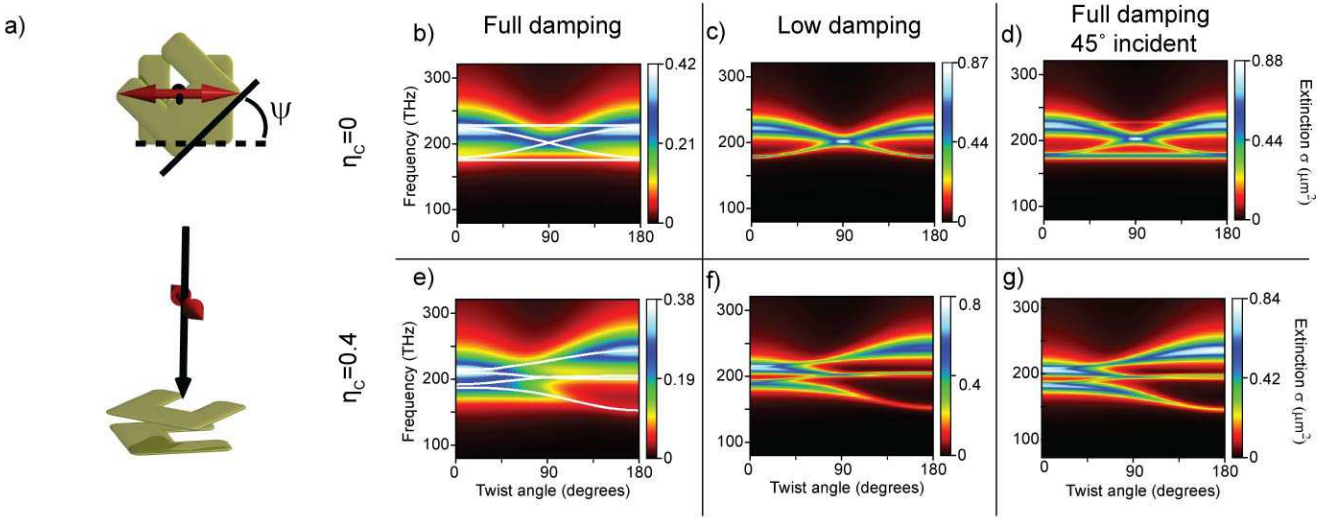}
\caption{Extinction cross sections $\sigma_{ext}$ versus frequency
and twist angle for an SRR stereodimer structure. Panel (a) shows
the geometry (top view and side view) in which two SRRs are
vertically stacked. The upper SRR is rotated around the $z$-axis by
the twist angle $\psi$. We calculate extinction for light impinging
from the $z$-direction with polarization along $x$, i.e., along the
base of the lower SRR in (b,c,e,f).  In (d,g) we use $45^\circ$
incidence in the $xz$-plane, so that the H-field of the excitation
light directly couples also to the magnetic polarizability. Panels
(b), (c) and (d) show extinction assuming no cross coupling term
($\eta_C=0$) while (e), (f) and (g) show extinction assuming strong
magneto-electric coupling $\eta_C=0.4$). Panels (b) and (e) assume
the damping rate of gold $\gamma=1.25\times10^{14}~$s$^{-1}$. To
more clearly bring out the four mode structure, we reduce the
damping ten-fold for the calculations in (c,d,f, g). There are four
modes present in the system. White lines in (b,e) indicate the
frequencies of the modes, as taken from the resonances in the
low-damping case, i.e., the resonances in panels (d,g).  Since
}\label{Fig3:stereosystem}
\end{figure*}

\section{A coupled system: Split ring dimers\label{section:coupledSRR}}
So far, this manuscript has focused purely on the scattering
properties of single magneto-electric point scatterers. In the
remainder of the paper we illustrate that our method can be easily
used to analyze multiple scattering by magneto-electric scattering
clusters.
 In order to calculate
the response of a system of coupled magneto-electric dipoles, we generalize the
general self-consistent equation that describes scattering of
clusters of electric dipoles $\bm{p}$ as reviewed in~\cite{Abajo07}.
Assuming a system of $N$ magneto-electric point scatterers situated
at positions $\bm{r}_1\ldots\bm{r}_{N}$, the response upon illumination
by an incident field $(\bm{E}_{\mathrm{in}}(\mathbf{r}),\bm{H}_{\mathrm{in}}(\mathbf{r}))$
is determined by a set of $N$ self consistent equations for the
induced dipole moments in each scatterer. The dipole moment induced
in scatterer $n$ with polarizability tensor $\boldsymbol{\alpha}_n$ is
\begin{equation} \left(
\begin{array} {c}
\bm{p}_n \\
\bm{m}_n
\end{array} \right)
=\boldsymbol{\alpha}_n\left[\left(
\begin{array} {c}
\bm{E}_{\mathrm{in}}(\mathbf{r}_n)\\
\bm{H}_{\mathrm{in}}(\mathbf{r}_n)
\end{array} \right)
+\sum_{\substack{q=1\ldots N \\ q\neq n}}
\boldsymbol{G }^0(\bm{r}_n,\bm{r}_q) \left(
\begin{array} {c}
\bm{p}_q\\
\bm{m}_q
\end{array}
\right)\right]
\label{Eq:selfconsistent}
\end{equation}
Using this equation we can attempt to reinterpret recent measurements that evidence significant coupling in split rings in 2D arrays, as well as in oligomers~\cite{Sersic09,Giessen09,Feth10,guo07}.
Here we focus on the extinction
of a dimer of split rings in socalled `stereodimer'configuration,
first studied by Liu et al.~\cite{Giessen09}. Figure ~\ref{Fig3:stereosystem} shows such a `stereodimer', consisting of
two SRRs in vacuum ($V=200\times 200 \times 30$ nm$^3$, resonant at a wavelength around 1500~nm), both parallel to the $xy$ plane, vertically stacked with a
small height difference of 150~nm. The upper SRR is rotated by a
twist angle $\psi$ around the $z$-axis. On the basis of the report
by Liu et al.~\cite{Giessen09}, we expect two resonance peaks with
an angle dependent splitting, which can be explained in an LC model
as the summed  effect of electric dipole-dipole
coupling and magnetic dipole-dipole coupling.

We calculate the extinction versus twist angle and wavelength of an
incident beam incident from the $+z$ direction, with
$x$-polarization. This beam directly excites $\bm{p}_x$ in both rings, which also drive each other. We first
analyze the experiment assuming that there is no magneto-electric
coupling term (setting $\eta_C=$0, although we keep $\eta_E=$0.7 and
$\eta_H=0.3$). As
Fig.~\ref{Fig3:stereosystem}(b) shows, the extinction shows a
single strong resonance that is blueshifted relative to the single
SRR resonance at 200~THz. As a function of twist angle, this broad
resonance redshifts to 200~THz at a twist of 90$^\circ$, and shifts
back to 220~THz at a twist of 180$^\circ$. There is no sign of a
second resonance, which might be hidden below the strong
resonance. To bring out the second resonance more clearly, we reduce
the loss in
Fig.~\ref{Fig3:stereosystem}(b), to a 10 times lower value  $\gamma=1.25\cdot 10^{13}~$s$^{-1}$) for gold in
Fig. (c) and (d). For this almost absorption-free system,
Fig.~\ref{Fig3:stereosystem}(c) indeed shows two resonances in
extinction. The blue shifted resonance is now observed to cross with
a narrow red shifted resonance. The crossing is symmetric around $90^\circ$
and is consistent with the hybridization of an electric dipole fixed along $x$, with a second one above it twisted by
an amount $\psi$. The two branches have a very different width and strength,
consistent with the fact that a symmetric configuration of dipoles
couples more strongly to external fields (blue shifted resonance), than an antisymmetric
`dark' configuration (red shifted resonance).

 To verify whether
the two resonances observed in Fig.~\ref{Fig3:stereosystem}(a) are
  all resonances in the system, we change the angle of
incidence to 45$^\circ$ in the $xz$ plane, so that the exciting
field has an $H_z$ component to drive the split rings, in addition
to an $E_x$ component. Figure~\ref{Fig3:stereosystem}(d) shows that
in this case four resonances occur in extinction. In addition to the
two curved bands excited by $E_x$, there are also two non-dispersive
bands with a twist independent splitting. Obviously, these bands are
due to the coupling of two magnetic dipoles in symmetric (broad and
intense band) and antisymmetric head-to-tail configuration. The
existence of four instead of two modes is a new insight compared to
LC circuit models~\cite{Giessen09,Giessen10}, but is logical in view
of the fact that split rings have both a magnetic and an electric
response, which are decoupled under the assumption $\eta_C=0$.

Next we analyze the extinction in presence of magneto-electric
coupling, setting $\eta_C= 0.4$. Again, we first examine the
extinction in presence of realistic loss ($\gamma=1.25\cdot
10^{14}~$s$^{-1}$) for gold in Fig.~\ref{Fig3:stereosystem}(e). As
also predicted by FDTD simulations by Liu et al.~\cite{Giessen09},
there appear to be two bands. The blue-shifted band is again  very
broad, but now has  a frequency shift away from the single SRR
resonance that is significantly larger for twist angle $180^\circ$
than for $0^\circ$. These effects were explained by Liu et al. as
due to an additive (subtractive) correction to the dominant electric
hybridization at twist angle $180^\circ$ ($0^\circ$) that occurs due
to magnetic dipole coupling.  A surprise is that  the diagram is not
symmetric anymore around $90^\circ$ twist as in the case of zero
magnetic coupling. Instead, the extinction appears to show an
anticrossing at twist angle $60^\circ$ These features were also
predicted by FDTD simulations by Liu et al.~\cite{Giessen09}
However,  the presence of an anticrossing at twist angle
$\psi=60^\circ$ could not be interpreted Liu et al ~\cite{Giessen09}
within an LC electrostatic circuit model, except by invoking higher
order multipolar corrections. Here we see that a purely dipolar
model may also explain all features of the experiment provided that
magneto-electric coupling is accounted for. 
While we do not claim that multipolar effects are not present in
actual experiments, it is an important insight that split ring
polarizabilities with magneto-electric coupling terms may provide
much richer physics then expected from electrostatic circuit theory.
A main advantage of
point dipole theory is that the underlying mode structure does not
need to be recouped from FDTD simulations, but is easily resolved by
repeating a calculation of extinction cross sections with low loss
(as done in Fig.~\ref{Fig3:stereosystem}), or by analyzing the
poles of the coupling matrix in Eq.~(\ref{Eq:selfconsistent}) that
relates $(\bm{p},\bm{m})$ to
$(\bm{E}_{\mathrm{in}},\bm{H}_\mathrm{in})$. The computational
effort for $N$ split rings is equivalent to diagonalizing or
inverting a $6N\times 6N$ matrix.

To more clearly bring out all the resonances we artificially reduce
the damping $\gamma=1.25\cdot 10^{13}~$s$^{-1}$ to ten times less
than the damping of gold, and plot the response of the system under
normal incidence (f) and $45^\circ$ incidence (g) in
Fig.~\ref{Fig3:stereosystem} (f,g). The anticrossing at twist
angle $\psi=60^\circ$ appears to be due to the coupling of four
modes, as opposed to the intuition from LC circuit theory that only
two resonances anticross.
The existence of four, rather than two modes in a split ring dimer
appears surprising and is a second indication of the rich physics of magneto-electric scatterers. Intuition from LC circuits is that
although the subspace of driving fields is two dimensional ($E_x$
and $H_z$), nonetheless only one mode per split ring exists. The
usual reasoning in LC models is that the relation between electric
and magnetic dipole moment is completely fixed and independent of
driving, since the loop current and accumulated charge are directly
related.  Such a constraint is not general: in electrodynamic
multipole expansions, magnetic polarizabilties are determined
independently from the electric ones. The intuition from LC theory
that there is only one mode per scatterer is only retrieved in our
model right at the limit of strongest magneto-electric coupling
$\eta_C=\sqrt{\eta_E\eta_H}$, since in that case one polarizability
is identically zero. We   note that the values $\eta_E=0.7,
\eta_H=0.3, \eta_C=0.4$ used in this work (that we fitted to our
angle-resolved transmission experiments on 200x200 nm Au split rings
on glass) are close to the limit of strong magneto-electric
coupling. Whether a general argument exists why physical scatterers
are or are not exactly at the limit of strongest magneto-electric
coupling $\eta_C=\sqrt{\eta_E\eta_H}$ is a question outside the scope
of this paper.

\section{Conclusion}
In conclusion, we have developed a new multiple scattering theory by
means of which we can calculate scattering and extinction for any
magneto-electric scatterer with known polarizability tensor, as well
as for arbitrary finite clusters. As opposed to LC circuit models,
our new  model obeys energy conservation,  contains all interference
effects, and allows quantitative prediction of absolute cross
sections, spectral linewidths and lineshapes. While outside the
scope of this paper, the theory is readily extended to deal with
arbitrary periodic lattices by generalizing Ewald lattice
sums~\cite{Abajo07} to deal with both $\bm{E}$ and $\bm{H}$.   Since
the electrodynamic polarizability tensor can be directly constructed
from electrostatic circuit theory, we expect that our model is
readily applicable to many current experiments using chiral and
nonchiral metamaterial building blocks for which electrostatic
models have been proposed.

Our model does not give any insight into whether the response of a
given structure is truly dipolar or not. Also, our model  does not
provide any insight or quantitative predictions based on microscopic
considerations for the magnitude of the polarizability. For such
microscopic considerations, based on, e.g., current density
distributions derived from full wave simulations, we refer to
~\cite{Rockstuhl06b,Rockstuhl07,Zhou07,Corrigan08,Pors10,RockstuhlMulti}.
Rather, our model allows one to verify if specific data or
microscopic calculations are consistent at all with point dipole
interactions, allowing to verify or falsify common intuitive
explanations in literature that have sofar always been based on
electrostatic considerations. Also, our model allows one to assess
if a single polarizability tensor indeed can describe a range of
different experiments with, e.g., split ring clusters, as should be
expected from a consistent model. Finally, our model is the simplest
electrodynamical model to consistently describe how metamaterials
and photonic crystals are formed from magneto-electric scatterers. A
first step is to confirm the parameters used in this work for
$\eta_E, \eta_H$ and $\eta_C$ by targeted experiments. While the
value for $\eta_E$ used in this work is consistent with the
extinction cross section measured by Husnik et al.~\cite{Husnik08},
we propose that the new insight that magneto-electric coupling is
far stronger than the magnetic polarizability be confirmed by
off-normal circularly polarized extinction measurements as proposed
in section~\ref{section:singleSRRexp}.

The most important property of our theory is that a polarizability
tensor validated for a single scatterer can readily be used to
predict all quantitative scattering properties of composite lattices
and antennas. We hence expect that new insights can be obtained in
effective medium constants of metamaterial arrays. Our analytical
model not only facilitates design, but will also for the first time
allow to determine rigorously whether, even in the ideal case (no
loss, no multipole corrections), metamaterial building blocks can
give rise to a desired $\epsilon$ and $\mu$, despite the large
importance of electrodynamic
corrections~\cite{RockstuhlWAT,Rockstuhl06,Sersic09}. In addition to
generating new insights for metamaterials, our theory also opens new
design routes for gratings and antennas with unprecedented
polarization properties. As an example, in this paper we analyzed
the four mode anticrossing due to magneto-electric coupling in
stereo-dimers. This analysis is easily extended to magneto-electric
Yagi-Uda antennas, diffractive gratings of chiral building blocks,
and magneto-inductive waveguides that may provide new ways  to
control the propagation and emission of
light~\cite{Koenderink06,Koenderink09,Genovguide}.

\begin{acknowledgments}
We thank Ad Lagendijk for stimulating and inspirative insights, as well as Dries van Oosten and Lutz Langguth for discussions. This work is part of the research program
of the ``Stichting voor Fundamenteel Onderzoek der Materie
(FOM),'' which is financially supported by the ``Nederlandse
Organisatie voor Wetenschappelijk Onderzoek (NWO).''
\end{acknowledgments}

\appendix
\section{Unit system\label{appendix}}
Throughout this paper we used  units that significantly simplify
notation throughout, as they maximimize the interchangeability of
electric and magnetic fields. Conversion to SI units is summarized
in Table~\ref{table}.
\begin{table}
\begin{tabular}{|l||c|l|}
\multicolumn{3}{c}{}\\
  Quantity & Symbol & Relation to SI \\
\hline
  Electric field & $\bm{E}$ &  $\bm{E}_{\mathrm{SI}}$ \\
  Magnetic field & $\bm{H}$ & $ Z \bm{H}_{\mathrm{SI}}$ \\
  &   &   \\
Electric dipole moment  &  $\bm{p}$ &  $ \bm{p}_{\mathrm{SI}}/(4\pi\epsilon)$ \\
Magnetic dipole moment  &  $\bm{m}$ &  $ \bm{m}_{\mathrm{SI}} (Z/(4\pi))$ \\
  &   &   \\
  Electric-electric polarizability & $\boldsymbol{\alpha}_{EE}$ &    $ \boldsymbol{\alpha}^{\mathrm{SI}}_{EE}/(4\pi\epsilon)$ \\
  Magnetic-magnetic polarizability  & $\boldsymbol{\alpha}_{HH}$ &    $\boldsymbol{\alpha}^{\mathrm{SI}}_{HH}/(4\pi)$ \\
  Electric-magnetic polarizability & $\boldsymbol{\alpha}_{EH}$ &   $\boldsymbol{\alpha}^{\mathrm{SI}}_{EH}(c/(4\pi))$\\
  Magnetic-electric polarizability & $\boldsymbol{\alpha}_{HE}$  &  $\boldsymbol{\alpha}^{\mathrm{SI}}_{HE}(Z/(4\pi))$   \\
  &   &   \\
Electric-electric Green tensor & $\boldsymbol{G}_{EE}$ &    $ 4\pi\epsilon \boldsymbol{G}^{\mathrm{SI}}_{EE}$\\
Magnetic-magnetic Green tensor & $\boldsymbol{G}_{HH}$ &    $ 4\pi \boldsymbol{G}^{\mathrm{SI}}_{EE}$\\
Electric-electric Green tensor & $\boldsymbol{G}_{EH}$ &    $ 4\pi/Z\, \boldsymbol{G}^{\mathrm{SI}}_{EE} $\\
Magnetic-magnetic Green tensor & $\boldsymbol{G}_{HE}$ &    $ 4\pi/c\, \boldsymbol{G}^{\mathrm{SI}}_{EE}$\\
\multicolumn{3}{c}{}\\
\end{tabular}\\
\caption{Conversion between SI units and the unit system used
throughout this paper.}\label{table}
\end{table}
For the conversion in Table~\ref{table}, we use $\epsilon$ for the
host dielectric constant, $c$ for the velocity of light, and $Z$ for
the impedance of the background medium. In this unit system, a plane
wave has $|\bm{E}|/|\bm{H}|=1$, and intensity $I=|\bm{E}|^2/(2Z)$,
since the Poynting vector is $\bm{S}=1/(2Z)
\mathrm{Re}(\bm{E}^{*}\times\bm{H})$. In these units, the
cycle-averaged work done by an electric field $\bm{E}$ to drive an
oscillating $\bm{p}$ equals $W=2\pi k/Z \mathrm{Im}(\bm{E}\cdot
\bm{p})$. The magnetic counterpart is $W=2\pi k/Z
\mathrm{Im}(\bm{H}\cdot \bm{m})$


\begin{thebibliography}{}
\bibitem{Veselago68}
V. G. Veselago, Sov.Phys. USPEKHI \textbf{10}, 509-514 (1968).

\bibitem{Pendry00}
J. B. Pendry, Phys. Rev. Lett. \textbf{85}, 3966 (2000).

\bibitem{Pendry01}
J. B. Pendry, Physics World \textbf{14}, 47 (2001); C. M.
Soukoulis, S. Linden, and M. Wegener, Science \textbf{315}, 47
(2007); V. M. Shalaev, Nature Photonics \textbf{1}, 41 (2007).


\bibitem{Pendry06} 
U. Leonhardt, Science \textbf{312},1777 (2006); J. B. Pendry, D.
Schurig, and D. R. Smith, \emph{ibid}, 1780 (2006).



\bibitem{Smith00}
D. R. Smith, W. J. Padilla, D. C. Vier, S. C. Nemat-Nasser, and S.
Schultz,  Phys. Rev. Lett. \textbf{84}, 4184  (2000); W. J. Padilla,
A. J. Taylor, C. Highstrete, M. Lee, and R. D. Averitt,
\emph{ibid.}~\textbf{96}, 107401 (2006); S. Linden, C. Enkrich, M.
Wegener, J. Zhou, T. Koschny, and C. M. Soukoulis, Science
\textbf{306}, 1351 (2004).


\bibitem{Enkrich05}
C. Enkrich, M. Wegener, S. Linden, S. Burger, L. Zschiedrich, F.
Schmidt, J. F. Zhou, T. Koschny, and  C. M. Soukoulis, Phys. Rev.
Lett. \textbf{95},  203901 (2005).


\bibitem{Rockstuhl06}
C. Rockstuhl, T. Zentgraf, H. Guo, N. Liu, C. Etrich, I. Loa, K.
Syassen, J. Kuhl, F. Lederer, and H. Giessen, Appl. Phys. B
\textbf{84}, 219 (2006).

\bibitem{Klein06}
M. W. Klein, C. Enkrich, M. Wegener, C. M. Soukoulis, and S. Linden,
Opt. Lett. \textbf{31}, 1259 (2006).

\bibitem{Sersic09}
I. Sersic, M. Frimmer, E. Verhagen and A. F. Koenderink, Phys. Rev.
Lett. \textbf{103}, 213902 (2009).

\bibitem{Lahiri10} B. Lahiri, S. G. McMeekin, A. Z. Khokhar, R. M.
De La Rue, and N. P. Johnson, Opt. Expr. \textbf{18}, 3210 (2010).   






\bibitem{Shalaev05}V. M. Shalaev, W. Cai, U. K. Chettiar, H.-K. Yuan, A. K.
Sarychev, V. P. Drachev, and A. V. Kildishev, Opt. Lett.
\textbf{30}, 3356 (2005).

\bibitem{Dolling05}
G. Dolling, C. Enkrich, M. Wegener, J. F. Zhou, C. M. Soukoulis, and
S. Linden, Opt. Lett. \textbf{30}, 23 (2005).

\bibitem{Dolling06}
G. Dolling, C. Enkrich, M. Wegener, C. M. Soukoulis, and S. Linden,
Opt. Lett. \textbf{31}, 12 (2006).

\bibitem{Dolling07}
G. Dolling, C. Enkrich, M. Wegener, C. M. Soukoulis, and S. Linden,
Science \textbf{312}, 892 (2007)

\bibitem{Valentine08}
J. Valentine, S. Zhang, T. Zentgraf, E. Ulin-Avila, D. A. Genov, G.
Bartal, and X. Zhang, Nature \textbf{455}, 376 (2008).

\bibitem{Waele10}
S. P. Burgos, R. de Waele, A. Polman,  and H. A. Atwater, Nature
Mat. \textbf{9}, 407 (2010).

\bibitem{Husnik08}
M. Husnik, M. W. Klein, N. Feth, M. K\"{o}nig, J. Niegemann, K.
Busch, S. Linden and M. Wegener, Nature Photonics \textbf{2}, 614
(2008).

\bibitem{Rockstuhl06b}
 C. Rockstuhl, F. Lederer, C. Etrich, T. Zentgraf, J. Kuhl, and
H. Giessen, 
Opt. Expr. \textbf{14}, 8827 (2006).

\bibitem{Corrigan08} T. D. Corrigan, P. W. Kolb, A. B. Sushkov, H. D. Drew, D. C. Schmadel, and R. J.
Phaneuf, Opt. Expr. \textbf{16}, 19850 (2008).

\bibitem{Pors10} A. Pors, M. Willatzen, O. Albrektsen, and S. I.
Bozhevolnyi,  J. Opt. Soc. Am. B \textbf{27}, 1680  (2010).  





\bibitem{NordlanderScience} E. Prodan, C. Radloff, N. J. Halas and P.
Nordlander, Science \textbf{302}, 419 (2003).




\bibitem{Banzer10} P. Banzer, U. Peschel, S. Quabis, and G. Leuchs,
Opt. Expr. \textbf{18}, 10905 (2010).   

\bibitem{Feth10}
N. Feth, M. K\"{o}nig, M. Husnik, K. Stannigel, J. Niegemann, K.
Busch, M. Wegener, and S. Linden, Opt. Express \textbf{18}, 6545
(2010).



\bibitem{decker09} M. Decker, S. Burger, S. Linden, and M. Wegener,
Phys. Rev. B \textbf{80}, 193102 (2009).


\bibitem{Gansel09}
J. K. Gansel, M. Thiel, M. S. Rill, M. Decker, K. Bade, V. Saile,
G. von Freymann, S. Linden, and M. Wegener, Science \textbf{325},
1513 (2009).
\bibitem{Plum09}
E. Plum, J. Zhou, J. Dong, V. A. Fedotov, T. Koschny, C. M.
Soukoulis, and N. I. Zheludev, Phys. Rev. B \textbf{79}, 035407
(2009).
\bibitem{Plum09b}
E. Plum, X.-X. Liu, V. A. Fedotov, Y. Chen, D. P. Tsai, and N. I.
Zheludev, Phys. Rev. Lett. \textbf{102}, 113902 (2009).

\bibitem{Zhang09}
S. Zhang, Y.-S. Park, J. Li, X. Lu, W. Zhang, and X. Zhang, Phys.
Rev. Lett. \textbf{102}, 023901 (2009).

\bibitem{Wang09}
B. Wang, J. Zhou, T. Koschny, M. Kafesaki, and C. M. Soukoulis, J.
Opt. A: Pure Appl. Opt \textbf{11}, 114003 (2009).


\bibitem{plum07} E. Plum, V. A. Fedotov and N. I. Zheludev, J.
Opt. A: Pure Appl. Opt. \textbf{11},074009:1-7 (2009).



\bibitem{decker07} M. Decker, M. W. Klein, M. Wegener, S. Linden,
   Opt. Lett. \textbf{32}, 856 (2007).  


\bibitem{decker10} M. Decker, R. Zhao, C. M. Soukoulis, S. Linden, M. Wegener,
Opt. Lett. \textbf{35},  1593 (2010). 

\bibitem{Giessen09}
N. Liu, H. Liu, S. Zhu and H. Giessen, Nature Photonics \textbf{3},
157 (2009).




\bibitem{guo07} H. Guo, N. Liu, L. Fu, T. P. Meyrath, T. Zentgraf, H. Schweizer,
and H. Giessen,    Opt. Expr. \textbf{15}, 12095 (2007).   

\bibitem{RockstuhlMulti} J. Petschulat, J. Yang, C. Menzel, C. Rockstuhl,
A. Chipouline, P. Lalanne, A. Tüennermann, F. Lederer, and T.
Pertsch, Opt. Express \textbf{18}, 14454 (2010).

\bibitem{Giessen10} H. Liu, J. X. Cao, S. N. Zhu, N. Liu, R. Ameling and H. Giessen,
Phys. Rev. B \textbf{81}, 241403(R) (2010).






\bibitem{Rockstuhl07} C. Rockstuhl, T. Zentgraf, E. Pshenay-Severin, J. Petschulat, A.
Chipouline, J. Kuhl, T. Pertsch, H. Giessen, and F. Lederer,  Opt.
Expr. \textbf{15}, 8871 (2007). 

\bibitem{Zhou07} J. Zhou,  Th. Koschny, and C. M Soukoulis, Opt.
Expr. \textbf{15}, 17881 (2007). 








\bibitem{Lagendijk96}
A. Lagendijk and B. A. van Tiggelen, Phys. Rep. \textbf{270}, 143
(1996).
\bibitem{Pedro98}
P. de Vries, D. V. van Coevorden and A. Lagendijk, Rev. Mod. Phys.
\textbf{70}, 2 (1998).
\bibitem{Abajo07}
F. J. Garc\'{i}a de Abajo, Rev. Mod. Phys. \textbf{79}, 1267 (2007).


\bibitem{Landau}
L. D. Landau and E. M. Lifshitz, \textit{Electrodynamics of
Continuous Media}, Pergamon, Oxford (1960).

\bibitem{Lindell_biisotropicbook}
I. V. Lindell, A. H. Sihvola, S. A. Tretyakov, and A. J. Viitanen,
\textit{Electromagnetic Waves in Chiral and Bi-Isotropic Media},
Artech House, Norwood MA (1994).

\bibitem{MerlinPNAS} R. Merlin, Proc. Nat. Acad. Sci. \textbf{106}, 1693 (2009).


\bibitem{Weber04}
W. H. Weber and G. W. Ford, Phys. Rev. B. \textbf{70}, 125429
(2004).

\bibitem{Koenderink06}
A. F. Koenderink and A. Polman, Phys. Rev. B \textbf{74}, 033402
(2006). (2007).


\bibitem{Garcia-Garcia05}
J. Garc\`{i}a-Garc\`{i}a, F. Mart\`{i}n, J. D. Baena, R. Marq\`{e}s
and L. Jelinek, J. Appl. Phys. \textbf{98}, 033103 (2005).


\bibitem{BohrenHuffman}
C. F. Bohren and D. R. Huffman, \textit{Absorption and Scattering of
Light by Small Particles}, John Wiley \& Sons, New York (1983).




\bibitem{APLSoukoulis}
N. Katsarakis, T. Koschny, M. Kafesaki, E. N. Economou, and C. M.
Soukoulis,  Appl. Phys. Lett. \textbf{84}, 2943 (2004).



\bibitem{tfootnote} Note that the above equation should strictly be written with $\boldsymbol{\alpha}$ replaced by the $t$-matrix, which
is directly proportional to the dynamic polarizability $\boldsymbol{\alpha}$ for point scatterers~\cite{Lagendijk96}.



%
%
%

\bibitem{jacksonbook} J. D. Jackson, \emph{Classical Electrodynamics (3rd ed.}, John Wiley \& Sons, New York (1999).


\bibitem{vdHulst}
H.~C. van~de Hulst, {\it Light Scattering by Small Particles\/}
(Dover, New York, 1981).




\bibitem{wokaun}  M. Meier and A. Wokaun, Opt. Lett. \textbf{8},
581 (1983); K. T. Carron, W. Fluhr, A. Wokaun and H. W. Lehmann,
J. Opt. Soc Am. B \textbf{3}, 420 (1986); K. L. Kelly, E.
Coronado, L. L. Zhao and G. C. Schatz, J. Phys. Chem. B
\textbf{107}, 668 (2003); A. Wokaun, J. P. Gordon and P. F. Liao, Phys. Rev. Lett. \textbf{48}, 1574 (1982).

\bibitem{Burresi09}
M. Burresi, D. van Oosten, T. Kampfrath, H. Schoenmaker, R. Heideman, A. Leinse and L. Kuipers,
Science \textbf{326}, 550 (2009); M. Burresi, T. Kampfrath, D. van Oosten, J. C. Prangsma, B. S. Song, S. Noda and L. Kuipers,
Phys. Rev. Lett. \textbf{105}, 123901 (2010).


\bibitem{silveirinha}
M. G. Silveirinha,Phys. Rev. B \textbf{82}, 085101 (2010)

\bibitem{soukouliscasimir} R. Zhao, J. Zhou, Th. Koschny, E. N. Economou, and C. M. Soukoulis, Phys. Rev. Lett. \textbf{103}, 103602 (2009);
M. G. Silveirinha and S. I. Maslovski, Phys. Rev. Lett. \textbf{105}, 189301 (2010);  R. Zhao, J. Zhou, Th. Koschny, E. N. Economou, and C. M. Soukoulis, Phys. Rev. Lett. \textbf{105}, 189302
 (2010).

\bibitem{Johnson} P. B. Johnson and R. W. Christy, Phys. Rev. B
\textbf{6}, 4370 (1972).


\bibitem{Sersictopublish} I. Sersic, A. Opheij and A. F.
Koenderink, \emph{in preparation\/}.

\bibitem{diagonalnote}Note that the diagonal elements of the
electrodynamic polarizability do contain contributions due to
off-diagonal elements in the electrostatic tensor. The radiation
damping correction in Eq.~(\ref{Eq:SRRradiationdamp}) mixes
$\eta_C$ onto the diagonal.

\bibitem{RockstuhlWAT} C. Menzel, T. Paul, C. Rockstuhl, T. Pertsch, S. Tretyakov, and F. Lederer,
Phys. Rev. B \textbf{81}, 035320 (2010).

\bibitem{Koenderink09} A. F. Koenderink, Nano Lett. \textbf{9}, 4228 (2009).

\bibitem{Genovguide} H. Liu, D. A. Genov, D. M. Wu, Y. M. Liu, J. M. Steele, C. Sun, S. N. Zhu, and X. Zhang,
Phys. Rev. Lett. \textbf{97}, 243902 (2006).




\end{thebibliography}
\end{document}